# COVID-19 Severity Classification on Chest X-ray Images

Aditi Sagar[1], Aman Swaraj[2], Karan Verma[1]

National Institute of Technology Delhi[1], Indian Institute of Technology, Roorkee[2]
202211001@nitdelhi.ac.in[1], aman_s@cs.iitr.ac.in[2], karanverma@nitdelhi.ac.in[1]

**Corresponding Author:**

Aman Swaraj
Ph.D. Research Scholar,
Indian Institute of Technology, Roorkee

Institute Address:
Indian Institute of Technology Roorkee
Roorkee, Uttarakhand
India - 247667

Permanent Address:
488- Solanipuram,
Roorkee- 247667
Uttarakhand, India

Tel: 91-7217842795, 91-7783860847
Email: aman_s@cs.iitr.ac.in; amanswaraj007@gmail.com

## ABSTRACT

Biomedical imaging analysis combined with artificial intelligence (AI) methods has proven to be quite useful in order to diagnose COVID-19. So far, various classification models have been used for diagnosing COVID-19. However, classification of patients based on their severity level is not yet analyzed. In this work, we classify covid images based on severity of the infection. First, we pre-process the X-ray images using median filter and histogram equalization. Enhanced X-ray images are then augmented using SMOTE technique for achieving a balanced dataset. Pre-trained Resnet50, VGG16 model and SVM classifier are then used respectively for feature extraction and classification. The result of the classification model confirms that compared with the alternatives, with chest X-Ray images, the ResNet-50 model produced remarkable classification results in terms of accuracy (95%), recall (0.94), F1-Score (0.92), and precision (0.91).

## 1. INTRODUCTION

Diseases related to lung abnormality are very common among humans and COVID-19 is the recent most reported virus of that kind. The first case of Coronavirus Disease was reported in Wuhan, China in December 2019. In a very quick span of time, it spread out globally, and due to its rapid transmission, the World Health-Organization (WHO) declared it a Pandemic on Mar 11, 2020. The original virus that was detected in the initial phases of the pandemic was called as SARS – COV – 2. However, in the times that followed, the virus mutated in several variants such as the 'Delta (Delta - B.1.617.2) and OMICRON (Omicron - B.1.1.529) variants. According to WHO, although Omicron is less severe than the Delta variant, but it is 4.5 times more infectious. The classification of these variants is based on how easily they can spread and how severe the symptoms are. Different waves of corona virus in India and the variant responsible for the wave are depicted in Table 1. Globally, as of 18 February 2022, there have been 418 million confirmed cases of COVID-19, including 5.8 million deaths, reported to W.H.O. Fig. 1 shows the total number of confirmed cases of COVID-19 disease in top three countries.

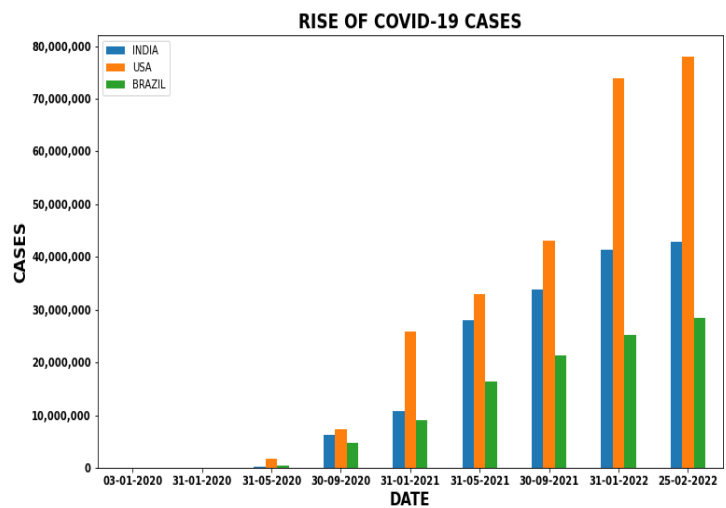

**Fig. 1: Confirmed cases of COVID-19 in top three countries from January 03, 2020 to February 25, 2022**

**TABLE 1: Different waves corresponding to different corona variants in India**

| Wave | Country | Corona virus variant |
|---|---|---|
| First wave, March 2020 | India | SARS – COV – 2 |
| Second wave, April 2021 | India | Delta – B.1.617.2 |
| Third wave, Nov 2021 | India | Omicron - B.1.1.529 |

Along with large scale loss of lives, it has also caused a significant pressure on the health care system and the economy in general. In our previous work, we came up with a hybrid forecasting model to predict future cases [1, 2]. However, a forecasting model can only give an overview of the rise of pandemic, but can't diagnose patients for that matter. Therefore, to quickly diagnose patients of covid-19 has become the pre-eminent need today in most healthcare facilities. In this connection, antigen tests such as RT-PCR tests have been sought out for detection of the virus, however reports have suggested they are only 60% accurate and therefore doctors have subsequently recommended X-Ray scans [3]. Furthermore, in a densely populated country like India, often the patients have to wait in a large queue to get medically diagnosed which overloads the clinicians and radiologists and negatively affects the patient's treatment and control of the pandemic.

Patients must be classified according to their severity levels since medical facilities including intensive care units (ICUs) and ventilators are inadequate to coping with this extremely infectious illness. Because the virus primarily affects the respiratory system, medical imaging can be used to detect the severity of the infection. In this connection, portable Chest X-rays (CXR) play an important role for imaging patients as they are informative and the equipment also remains disinfected even after repeated use. The severity of illness can be measured by chest x-rays in many of lung disorders [4]. Bilateral hazy opacity and airspace consolidation on Chest radiograph are hallmarks of COVID-19 lung infection [5]. ICU beds and mechanical ventilators are expected to be in limited supply at many hospitals. Chest X-ray could be useful in determining which patients should be put on mechanical ventilation, monitoring disease growth and treatment consequences while being on mechanical ventilation, and determining when it's safe to extubate.

In recent times, many deep learning and AI (Artificial Intelligence) techniques are being used for the diagnosis of many diseases in advancing medical research. Deep learning-based lung diagnosis can assist the radiologist in detecting COVID-19 signs in a patient quickly and precisely, as well as determining the severity stage in the patient. Over the last year, lots of new research articles have been published on COVID-19 and deep learning, with the majority of them focusing on disease detection rather than severity evaluation. However, Researchers have showed that assessing severity from X-ray radiographs is a powerful strategy for combating this extremely infectious disease. Deep learning approaches can be used to identify COVID patients who require significant clinical treatment becomes extremely important throughout the epidemic. The proper assessment of COVID-19 individuals at a preliminary phase is now a critical challenge if the rate of transmission as well as the mortality is to be reduced. With this motivation, we propose a methodology that can classify the patients based on their severity level.

We analyze multiple deep learning models and develop one that can accurately diagnose the stage of severity in COVID-19 patients using chest X-ray radiographs. We use publicly available dataset for Covid patients and aim to classify them into 3 categories as Normal, Non- Severe and Severe. The rest of the paper is organized as follows: section 2 is a literature review, and section 3 is a description of the methodology. Section 4 depicts all of the outcomes, and section 5 concludes with suggestions for further work.

## 2. LITERATURE REVIEW

To advance the biomedical research, artificial intelligence (AI) has transpired as foremost catalyst [6, 7]. Many deep learning and ML approaches incorporating biomedical image analysis have been employed in biomedical disease detection. These techniques can be widely used in the skin cancer classification [8], breast cancer detection [9] etc. Similarly, we may classify covid individuals with other people and assign them to different severity levels using chest x-ray images.

In general, COVID-19 diagnosis can be done in two categories using chest X-Ray images. The first is a categorization of COVID patients from healthy subjects or other patients of pneumonia or other lungs related infections and the second category is related to the severity evaluation of COVID patients.

X-ray radiographs and CT scans are often used to diagnose COVID-19 disease in several studies. In this connection, Ying et al. [10] proposed a model to distinguish COVID-19 patients from healthy persons and bacterial pneumonia patients from CT scans using the ResNet50 model, which had an accuracy of 86 %. Jin et al. [11] utilized CT images to develop a Convolutional neural network model that could differentiate 723 COVID-19 patients from 413 other patients, achieving 97.4 % sensitivity and 92.2 % specificity. Using CT scans, Xu et al. [12] distinguished 219 COVID-19 patients from 175 healthy persons and 224 influenza-A patients. He employed a convolutional neural network and attained an accuracy of 86.7 %.

Majority of the studies has concentrated on COVID-19 disease detection instead of COVID-19 severity evaluation [10-21]. Table 2 depicts some of the work in this area briefly.

**TABLE 2: Earlier works on medical images for diagnosis of Covid-19**

| Literature | Modality | Dataset | Task | Method | Result |
|---|---|---|---|---|---|
| Ying et al. [10] | CT SCAN | 88 COVID-19; 100 Bacterial Pneumonia; 86 Normal | **Classification:** Covid-19 / Bacterial Pneumonia / Normal | ResNet-50 | 86% |
| Jin et al. [11] | CT SCAN | 723 COVID-19; 413 Normal | **Classification:** Covid-19 / Normal | ResNet-50 | 97.4% (Sens.) 92.2% (Spec.) |
| Xu et al. [12] | CT SCAN | 219 COVID-19; 175 Normal; 224 influenza-A | **Classification:** Covid-19 / Normal / Influenza-A | ResNet-18 | 86.7% |
| Ghoshal et al.[13] | X-Ray | 170 COVID-19; Others(Not available) | **Classification:** Covid-19/Others | Bayesian CNN | 92.9% |
| Narin et al.[14] | X-Ray | 50 COVID-19; 50 Normal | **Classification:** Covid-19 /Normal | ResNet-50 | 96.1% |
| Zhang et al.[15] | X-Ray | 70 COVID-19; 1008 Others | **Classification:** Covid-19/Others | ResNet-18 | 96.0(Sens.) 70.7(Spec.) 0.952(AUC) |
| Wang et al.[16] | X-Ray | 45 COVID-19; 931 Bacterial Pneumonia; 660 Viral | **Classification:** Covid-19/Bacterial | Tailor-made CNN | 83.5% |

| Literature | Modality | Dataset | Task | Method | Result |
|---|---|---|---|---|---|
| | | Pneumonia; 1203 Normal | Pneumonia/Viral Pneumonia/ Normal | | |
| Hall LO et al.[17] | X-Ray | 102 COVID-19; 102 Pneumonia | **Classification:** Covid-19 / Pneumonia | ResNet-50, VGG16 | 89.2% |
| Narin A. Et al.[18] | X-Ray | 50 COVID-19; 50 Pneumonia | **Classification:** Covid-19 / Viral Pneumonia | ResNet-50 | 99.5% |
| Ozturk T et al.[19] | X-Ray | 127 COVID-19; 127 Normal | **Classification:** Covid-19 / Normal | Darknet | 98.08% |
| Emrah Irmak et al. [20] | X-Ray | 625 Covid-19; 625 Normal | **Classification:** Covid-19 / Normal | Tailor-made CNN | 99.20% |

However, several studies clearly emphasize on classifying the patients based on the severity of the infection. For instance, in [22], Chen et al. point out the significant difference in death rates between non-severe COVID-patients and severe COVID- patients, as well as the excess of attention they deserve. Zhu et al. [23] categorizes 131 chest X-ray images into different stages of severity with the help of transfer learning. Another study marked the importance of identifying primary or mild symptoms of COVID-19 patients which may eventually lead to severe and critical stages. Also, the average time taken for primary symptoms to develop into critical stage is only five days and therefore need urgent attention [24]. Although detecting critical symptoms in a patient is very much needed in the domain, only a handful of studies related to severity have been reported.

In this connection, Shan et al. [24] has used the concept of segmentation by segmenting the lung lobes area infected by covid disease. Additionally, they have developed a model that uses consolidation and ground-glass opacity to automatically rate COVID-19 disease severity. Similarly, According to Tang et al. [25], model can extract qualitative features from CT lung images and then can be used to discriminate between severe and non-severe illness. Authors have used only 176 CT images and after applying the model, they have achieved 87.5% accuracy. Xiao et al. [36] developed a residual network (ResNet34) for the severity staging in covid patients. Authors have used only 408 CT images and after applying the model, they have achieved 81.9% accuracy He et al. [27] further claimed that by using both severity assessment and lung segmentation, they could achieve better results. Table 3 highlights work done in this area.

**TABLE 3: Earlier worn on medical images for severity assessment of Covid-19**

| Literature | Modality | Dataset | Task | Method | Result |
|---|---|---|---|---|---|
| **Shan et al. [24]** | CT SCAN | 249 covid - 19 | Severity assessment | VB-Net | 72% Accuracy |
| **Tang et al.[25]** | CT SCAN | 176 covid - 19 | Severity assessment | RF | 87.5% (Accuracy) 93.3% (TPR) 74.5% (TNR) |
| **Xiao et al. [26]** | CT SCAN | 408 covid - 19 | Severity assessment | ResNet34 | 81.9% Accuracy |
| **He et al. [27]** | CT SCAN | 666 covid - 19 | Severity assessment | $M^2$UNet | 0.952 (Sensitivity) |
| **Carvalho et al.[28]** | CT SCAN | 229 covid - 19 | Severity assessment | ANN | 82% Accuracy |
| **Zhang et al.[29]** | CT SCAN | 661 covid - 19 | Severity assessment | U-Net | 91.6% Accuracy |

| Jocelyn zhu et al. [30] | X-ray | 131 covid - 19 | Severity assessment | VGG16 | 8.5% MAE |

It is quite evident from table 2 and 3 that work in the direction of severity detection has scope for improvement compared to task of mere covid detection which has achieved significant performance. Our proposed model will not just detect the covid infected patient but also classify them into severe and Non-severe cases.

## 3. METHODOLOGY

This chapter provides a detailed description about the data sources and different methodologies used for data pre-processing, feature extraction, classification and finally for performance metrics used for the classification model. Section 3.1 provides the information about dataset used and number of chest x-ray images for different categories.

### 3.1. Dataset Description

For our research work, we filter out severe and non-severe images from two different datasets [31, 32]. Both of these datasets are in themselves a collection of multiple sources in the first place. Since our work is focused on severity detection, we don't make use of ordinary pneumonia images. We also discard certain images having ECG leads, wires, pacemakers, etc. owing to the fact that neural networks might extract additional features from these apparent noisy elements. Since severe images were less in count compared to non-severe ones, we used smote data augmentation to handle the class imbalance. We describe the augmentation methodology in detail in section 3.3. The final distribution is presented in Table 4.

**TABLE 4: Final class distribution of dataset used in the study**

| CLASS | IMAGES |
|---|---|
| **Normal** | 584 |
| **Severe (COVID-19)** | 584 |
| **Non-Severe (COVID-19)** | 584 |

### 3.2. Radiologist Scoring

The widely used standard approach for classifying the lung disease severity is based on radiological scoring of lung opacity and consolidation parameters [33-36]. In this study, we have divided the lungs into 6 regions with the help of a medical professional[1]. In the X-Ray images, two lines divide the lungs horizontally. It results into each lung having 6 regions. Each region is then rated from 0 to 2 based on the lesions. It is rated 0 if there is no involvement, 1 if consolidations are less than 50% and 2 if consolidation is more than 50%. Therefore, the maximum score that any patient can have is 12. Accordingly, non-severe patients are rated from score 1 to 8 and severe patients with lesion score 9 to 12.

---

[1]*Dr. Abinash Mishra, MD, (Government Hospital; Odisha, India)*

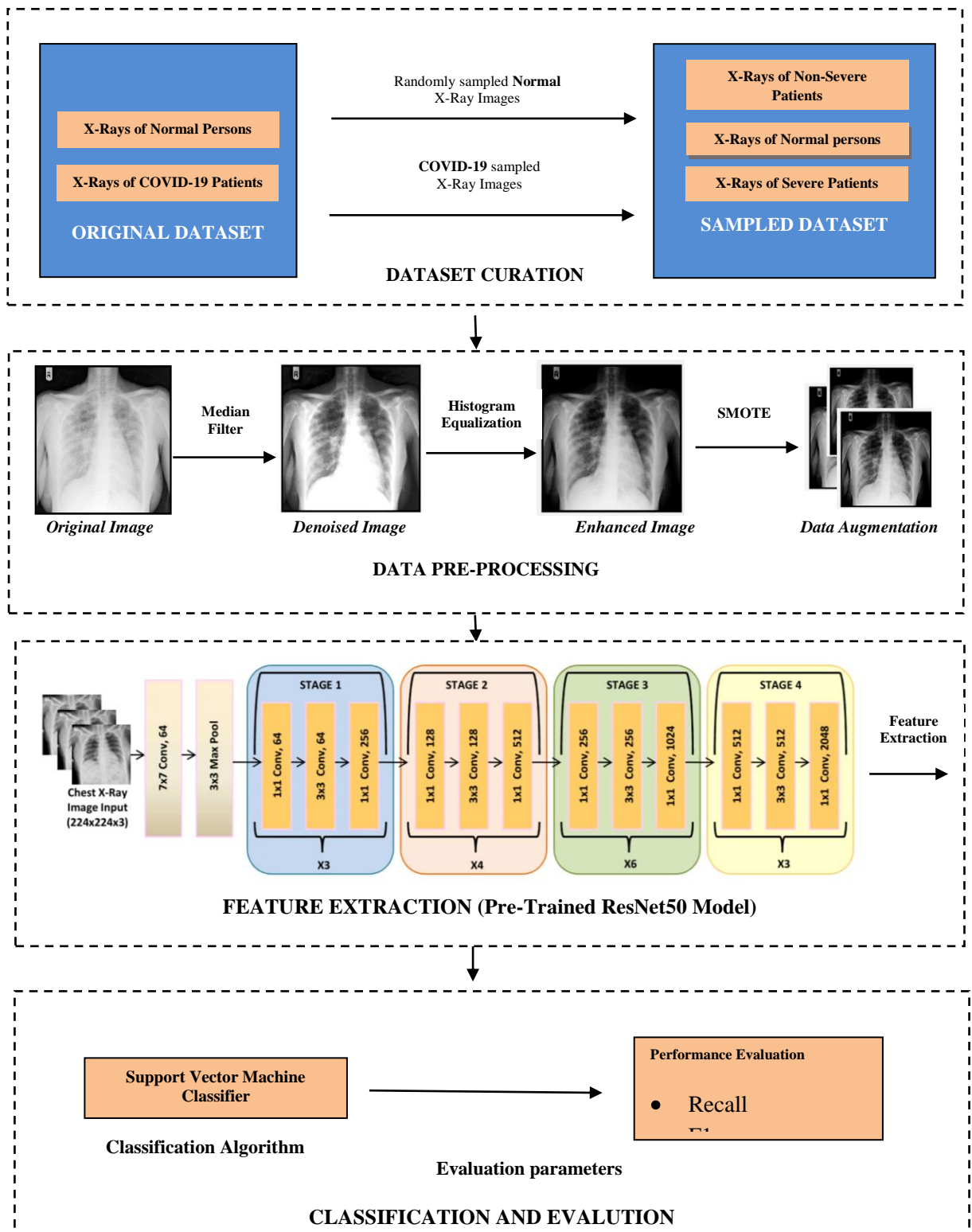

Fig. 2: Block Diagram of Proposed framework

### 3.3. Data Pre-Processing

Data pre-processing should be used before feeding the data into any machine learning model. The first step is to reshape all the images into one size because any neural network model receives input of the same size. Each sample image from the dataset is first reshaped to (224 x 224) to be used by the popular CNN models. Additionally, we have used one-hot encoding scheme for the labels of dataset where each class is converted into a numerical string. To enhance the images, histogram equalization technique is used over a median filter, and the final dataset is obtained. Fig. 2 depicts a block diagram of the proposed work.

*3.3.1. Image denoising and Enhancement*

Images of various resolutions, shapes, and sizes are included in the dataset. Furthermore, several of the images are of poor quality. The use of appropriate pre-processing techniques is required to perform image classification efficiently. We used the following image denoising techniques to eliminate the noise from the images:

a. Mean Filter,
b. Median Filter,
c. Bilateral Filter.

Among Mean, median and bilateral filter, Bilateral performed quite well, however due to its higher computation cost, we preferred median filter for our research work. Another reason for choosing median filters is its ability to preserve sharp edges as compared to the mean filter.

To achieve a filtered image, we have to insert the median of the values from input image at the center of the window in the output image. It may also be used to preserve an image's edges while decreasing random noise.
We have used the following image enhancing techniques to enhance the quality of the images.

a. Histogram Equalization,
b. Contrast Limited Adaptive Histogram Equalization (CLAHE).

CLAHE (Contrast Limited Adaptive Histogram Equalization) is a variant of histogram equalization that restricts contrast amplification in order to reduce noise amplification.

Histogram Equalization and CLAHE have demonstrated outstanding results in image denoising and enhancement. However, adding CLAHE raises the intensity of bones, which may have an impact on the classification model's efficiency since the neural network may interpret these (ribs and sternum bones) as the most important feature in determining COVID-19 severity. As a result, in our proposed approach, Histogram Equalization was chosen as the image enhancing technique.

*3.3.2. Data Augmentation*

Dealing with unbalanced datasets is one of the most prominent challenges in machine and deep learning. We have used data augmentation technique to handle our imbalance dataset. The reason of balancing the dataset is to achieve good accuracy for the minority class too.

We encounter several issues in real-world applications where we only have unequal data representations, with the minority class generally being the most significant. This issue poses a serious challenge to predictive modeling as under such situations, machine and deep learning algorithms can become biased towards the majority class as the model will learn only majority class features and will ignore the minority data points. To deal with this problem, we balanced our dataset using the Synthetic Minority Oversampling Technique (SMOTE).

SMOTE can be defined as an data augmentation techniques that enhances the prediction power of the class having less number of data points by producing the synthetic samples from the dataset, despite the fact that no data is lost. The working of SMOTE starts by choosing the nearby data points and connecting them with the help of a line and finally by plotting the new data or sample points on that line. In other words, a random data point from the minority class is first selected. It can be defined as follows-

$$S = x + u \cdot (x^R - x) \tag{1}$$

Where 'S' is sample space with $0 \leq u \leq 1$; $x^R$ is randomly chosen among nearest neighbors of x in minority class.

Then we use the KNN algorithm to select the k neighbors for the randomly selected data point. A randomly selected neighbor is chosen and a synthetic example is created at a randomly selected point between the two data points in feature space. Final distribution of the data is shown in table 4.

### 3.4. Model Description

We have used transfer learning for feature extraction. Features are extracted using ResNet50 model. Transfer learning is an approach in which the information retrieved by a CNN is transferred to address a distinct but comparable goal by training a model from scratch using different data, which is normally less in number [37, 38].

Feature extraction in a medical dataset is a tough process that must be approached with care. Because a radiograph comprises both low-level and high-level information such as edges, shape, texture, colour, blobs, and the ECG wires, and so on, choosing the right model is essential for separating desired features.

Kong et al. [39] discovered signs of a covid-19 infection in a patient. The formation of bilateral nodular, peripheral ground-glass opacities, and consolidators was the predominant sign of covid-19 in their study. A hazy lung opacity that looks thin on an X-ray photograph and is unable to cover any underlying bronchial walls or pulmonary arteries is known as ground-glass lung opacity. On the other side, having significant opacities that hide bronchial walls and capillaries is the polar opposite of consolidation.

In our research work, we employed VGG16 and ResNet50, two popular Convolutional Neural Network architectures. The best result is achieved by using the ResNet50 pre-trained model. We elaborate this in result section.

In general, the architecture of ResNet50 has 5 stages. Every ResNet architecture incorporates 7x7 and 3x3 kernel sizes for initial convolution and max-pooling, respectively. Stage 1 includes three Residual blocks. There are three layers in every block. Because of the inclusion of a shortcut connection between the input and output of each block, the convolutional blocks of a ResNet differ from those of regular CNNs. When employed as shortcut connection in ResNets [40], identity mappings can lead to greater optimization. As a result, deeper ResNets may be used, which are faster to train and less computationally costly.

The output of the last convolutional block of a 50-layer network (ResNet-50) [41] pre-trained on ImageNet is used to extract the features. Pre-trained deep networks' fully connected layers retain the fundamental structure of the entity included in the region of interest. When compared to lower layer features, deeper layer features encode class specific traits (such as shape, texture, and colour) and provide higher classification performance [42]. As a result, we use the output of the last layer to extract the features from the x-ray images. The Conv5 block produces a 7 x 7

x 2048 dimensional array, which is fed into the Support vector machine classifier. The same architecture is used in our proposed work till the last stage i.e. before fully connected layer. The final features are given to the classifier for severity classification.

## 3.5. CLASSIFICATION AND EVALUATION

Finally, we have used two classifiers named as Random Forest and Support Vector Machine. For multi-class classification of balanced dataset, the best result was achieved by support vector machine classifier with RBF kernel. Support vector machine classifier uses the hyperplane that can clearly classify the data points in N-dimensional feature space and in that feature space to find the regression line; we can use the Radial basis kernel function. The primary concept behind using kernel is that in higher dimensions, a linear classifier or regression curve becomes a non-linear classifier or regression curve in lower dimensions.
An RBF kernel has the following equation (Equation (2)):

$$F(X1, X2) = \exp(-gamma * \|X1 - X2\|^{\wedge 2}) \tag{2}$$

In the above equation, gamma specifies how much a single training point has on the other data points around it. **$\|X1 - X2\|$** is the dot product between the features.

Precision, recall, accuracy, and F1 score were among the performance metrics used to evaluate the proposed model's performance. True positive (TP), False Positive (FP), True Negative (TN), and False Negative (FN) metrics derived from the confusion metrics can be used to compute these metrics. These metrics can be used to evaluate the model.

- **Accuracy** is equal to the sum of total true positives and true negatives divided by the total values in the confusion matrix.

  $$\text{Accuracy} = (TP+TN)/(TP+FN+TN+FP) \tag{3}$$

- The difference between the number of true positive samples and the total number of positive samples is known as **precision**.

  $$\text{Precision} = (TP)/(TP+FP) \tag{4}$$

- **F1-score** is the harmonic mean of precision and recall.

  $$\text{F1 Score} = (2 * (\text{Precision} * \text{Recall}) / (\text{Precision} + \text{Recall})) \tag{5}$$

- The ratio of the total number of accurately categorized positive patients to the total number of positive patients is termed as **recall** or **sensitivity**.

  $$\text{Recall} = (TP)/(TP+FN) \tag{6}$$

## 4. Result & Comparison

First we start with pre-processing results in section 4.1. Section 4.2 provides the results of proposed model after applying the classifier. Finally, section 4.3 gives the comparative analysis of our work with respect to earlier works on severity detection in covid-19.

### 4.1. Pre-Processing Results

Median filter displayed better results as compared to others owing to its ability in preserving the edges and low computational cost. Using Histogram Equalization and Contrast Limited Adaptive Histogram Equalization to enhance images has yielded remarkable results. However, applying CLAHE increases the intensity of bones, which might affect the classification model's performance because the neural network may interpret these (ribs and sternum bones) as the most important feature in diagnosing COVID-19 severity.

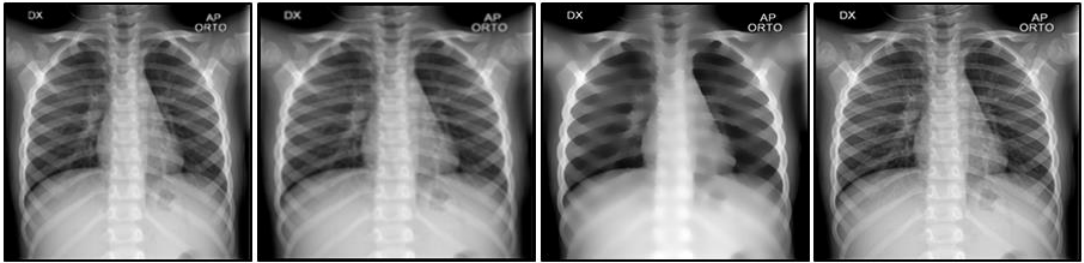

**Fig. 3: Various Image Denoising Techniques: (a) Original Image; (b) Median Filter; (c) Mean Filter, (d) Bilateral Filter**

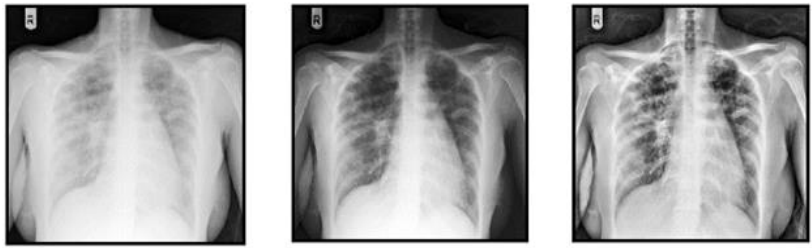

**Fig. 4: Various Image Enhancement Techniques: (a) Original Image; (b) Histogram Equalization; (c) CLAHE**

### 4.2. Result of proposed Methodology

Between the two models that is VGG16 and ResNet50 model, the highest accuracy was achieved by the ResNet50 model. Since, we are dealing with medical dataset, we not only considered accuracy but also the recall, precision and F1-score value. For model evaluation, we have drawn overlapped confusion matrix of the test dataset (Fig.5) with class labels as Non-severe, Severe and Normal. Table 5 summarizes the values of severity assessment metrics for the Normal, Severe, and Non-severe classes. We have achieved remarkable results with a total accuracy of 95% using ResNet50 model. Table 6 summarizes the comparison of the models employed in this research.

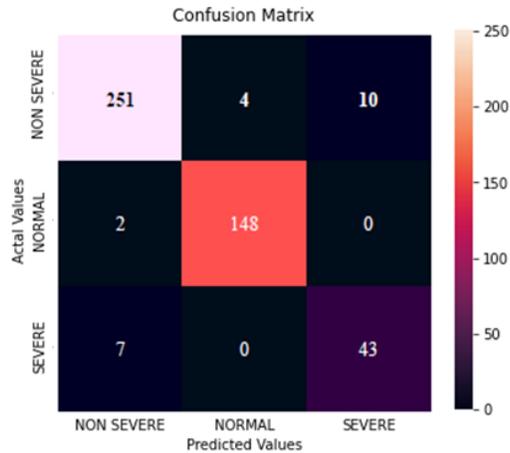

**Fig. 5: Overlapped Confusion Matrix for test dataset**

**Table 5: Support, precision, Recall, F1-score, and accuracy values of the proposed model for confusion matrix**

| CLASS | PRECISION | RECALL | F1 - SCORE | SUPPORT |
|---|---|---|---|---|
| **NORMAL** | 0.97 | 0.99 | 0.98 | 150 |
| **SEVERE** | 0.81 | 0.88 | 0.83 | 50 |
| **NON - SEVERE** | 0.97 | 0.95 | 0.96 | 265 |
| **AVERAGE** | 0.91 | 0.94 | 0.92 | 155 |
| **OVERALL ACCURACY** | | 95% | | |

**Table 6: Comparison of different models with different classifiers**

| MODEL | CLASSIFIER | |
|---|---|---|
| | Random Forest | SVM |
| **VGG** | 93% | 92% |
| **Resnet-50** | 92% | **95%** |

Finally, after using two commonly used deep learning model and classifiers, we can observe from the above table that the highest accuracy achieved by ResNet-50 model with SVM classier. Also, table 7 depicts how our model performs better compared to earlier works on severity detection in covid-19.

**Table 7: Comparison of our work with earlier works on severity detection in covid-19 X-ray images**

| Literature | Modality | Dataset | Task | Method | Result |
| --- | --- | --- | --- | --- | --- |
| **Shan et al. [24]** | CT SCAN | 249 covid - 19 | Severity assessment | VB-Net | 72% Accuracy |
| **Tang et al.[25]** | CT SCAN | 179 covid - 19 | Severity assessment | RF | 87.5% (Accuracy) 93.3% (TPR) 74.5% (TNR) |
| **Xiao et al. [26]** | CT SCAN | 408 covid - 19 | Severity assessment | ResNet34 | 81.9% Accuracy |
| **He et al. [27]** | CT SCAN | 666 covid - 19 | Severity assessment | $M^2$UNet | 0.952 (Sensitivity) |
| **Carvalho et al.[28]** | CT SCAN | 229 covid - 19 | Severity assessment | ANN | 82% Accuracy |
| **Zhang et al.[29]** | CT SCAN | 661 covid - 19 | Severity assessment | U-Net | 91.6% Accuracy |
| **Jocelyn zhu et al. [30]** | X-ray | 131 covid - 19 | Severity assessment | VGG16 | 8.5% MAE |
| **Our Work** | **X-Ray Images** | **1168 Covid-19** | **Severity assessment** | **ResNet50** | **95% Accuracy** |

## CONCLUSION

This work uses a multi-class classification deep learning model for determining the severity of COVID-19 illness. It includes two pre-trained deep learning models for the severity assessment: VGG16 and ResNet50. Between these two models we have observed that ResNet50 has performed better as compared to the VGG16 model. The diagnosis and severity of COVID-19 disease were determined using CXR photographs. Precision, recall, and F1-score were used to measure the prediction accuracy of both models. We also compared our outcomes to other existing methodologies to demonstrate the feasibility of our research. Using the ResNet50 model for feature extraction and a support vector machine classifier, we were able to attain an overall accuracy of 95%. If we consider the three waves of COVID-19, especially the second wave, we may infer that this approach will be very beneficial for doctors in classifying patients based on their severity levels and detecting COVID-19 disease in patients.

## Conflict of Interest / Competing Interests

The authors declare that they have no conflict of interest. Further, the authors have no relevant financial or non-financial interests to disclose. No funds, grants, or other support was received.

## ACKNOWLEDGMENTS

We thank Dr. Abinash Mishra, MD, (Government Hospital; Odisha, India) for classifying the lungs images as per their severity.